\documentclass{PoS}

\title{On Naturalness in type II seesaw models and the heavy Higgs masses}

\ShortTitle{On naturalness in type II seesaw and the heavy Higgs masses}

\author{\speaker{Mohamed Chabab} \\
LPHEA, Physics Department, FSSM,  Cadi Ayyad University, P.O.B. 2390, Marrakech, Morocco.\\ 
E-mail: \email{mchabab@uca.ac.ma}}
\author{{Michel Capdequi Peyranere}\\
LUPM, Universit\'e de Montpellier, F-34095 Montpellier, France.\\
E-mail: \email{michel.capdequi-peyranere@umontpellier.fr}}
\author{{Larbi Rahili} \\
LPHEA, Physics Department, FSSM,  Cadi Ayyad University, P.O.B. 2390, Marrakech, Morocco.\\
LMTI, FS, Ibn Zohr University, Agadir, Morocco,\\
E-mail: \email{rahililarbi@gmail.com}}

\abstract{Here we study naturalness problem in a type II seesaw model, and show that the Veltman condition modified (mVC) by virtue of the additional scalar charged states is satisfied at one loop within a region of the allowed parameter space of the model. The latter is severely constrained by unitarity and boundedness from below as well as  by consistency with the diphoton Higgs decay data of the LHC run 1. In this context, our analysis of the naturalness condition affects dramatically the masses of heavy Higgs bosons $H^0$, $A^0$, $H^\pm$ and $H^{\pm\pm}$, reducing the ranges of variation of $m_{H^\pm}$ and $m_{H^{\pm\pm}}$ with an upper bounds at 288 and 351 GeV, respectively, while the neutral Higgs bosons $H^0$, $A^0$ are found to be almost degenerate at about 207 GeV.}

\FullConference{38th International Conference on High Energy Physics\\
		3-10 August 2016\\
		Chicago, USA}

\begin{document}
\section{Introduction}
The discovery of the Higgs boson at the LHC definitely confirmed the mass generation mechanism through the spontaneous electroweak symmetry breaking. This discovery along with many experimental data of various experiments (LHC, Tevatron,.....)  reinforces the Standard Model (SM) as a successful theory in describing Nature. There are however many serious theoretical issues awaiting  resolution such as  the Naturalness problem and the neutrino masses. Hence one needs to modify the SM to tackle these problems and test the Higgs sector of the possible extension at colliders.  

In this work \cite{mc} we addressed the naturalness problem in the context of Type
II Seesaw model, dubbed HTM, with emphasis on its effect on the HTM parameter space.
More precisely, we studied how to soften the divergencies and how to gain some insight on
the allowed masses of the heavy scalars in the Higgs sector. 
Our study used the most general renormalisable Higgs potential of HTM and is essentially based on
dimensional regularisation approach which complies with unitarity and Lorentz invariance. Also, the
phenomenological analysis takes into account the full set of theoretical constraints, including 
unitarity and the consistent conditions of boundedness from below.
\section{Type II Seesaw Model and Higgs Bosons Masses}
Seesaw mechanism is implemented  in the Standard Model via addition of a scalar field $\Delta$ with hypercharge $Y=2$. The potential of this model is \cite{aa11,Perez08}
\begin{eqnarray}
\mathcal{L} &=&
(D_\mu{H})^\dagger(D^\mu{H})+Tr(D_\mu{\Delta})^\dagger(D^\mu{\Delta}) -V(H, \Delta) + \mathcal{L}_{\rm Yukawa}
\label{eq:DTHM}
\end{eqnarray}
where :
\begin{eqnarray}
V(H, \Delta) &=& -m_H^2{H^\dagger{H}}+\frac{\lambda}{4}(H^\dagger{H})^2+M_\Delta^2Tr(\Delta^{\dagger}{\Delta}) +[\mu(H^T{i}\sigma^2\Delta^{\dagger}H)+{\rm h.c.}] \nonumber \\
&& + \lambda_1(H^\dagger{H})Tr(\Delta^{\dagger}{\Delta}) + \lambda_2(Tr\Delta^{\dagger}{\Delta})^2
+ \lambda_3Tr(\Delta^{\dagger}{\Delta})^2 + \lambda_4{H^\dagger\Delta\Delta^{\dagger}H}
\label{eq:Vpot}
\end{eqnarray}
Triplet $\Delta$ and doublet Higgs $H$  
are represented by:
\begin{eqnarray}
\Delta &=\left(
\begin{array}{cc}
\delta^+/\sqrt{2} & \delta^{++} \\
\delta^0 & -\delta^+/\sqrt{2}\\
\end{array}
\right) \,\,\, {\rm and} \,\,\, H=\left(
                    \begin{array}{c}
                      \phi^+ \\
                      \phi^0 \\
                    \end{array}
                  \right)
                  \label{HDrep}\nonumber
\end{eqnarray}
After the spontaneous electroweak symmetry breaking, the Higgs doublet and triplet fields acquire their
vacuum expectation values $v_d$ and $v_t$ respectively,
\begin{eqnarray}
\langle \Delta \rangle&=\left(
\begin{array}{cc}
0 & 0 \\
\frac{v_t}{\sqrt{2}} & 0 \\
\end{array}
\right) \,\,\, {\rm and} \,\,\, \langle H \rangle=\left(
                    \begin{array}{c}
                      0 \\
                      \frac{v_d}{\sqrt{2}} \\
                    \end{array}
                  \right)
                  \label{HDrep}\nonumber
\end{eqnarray}
Extended Higgs sector: two $CP_{even}$ neutral scalars ($h^0$, $H^0$), one neutral pseudo-scalar $A^0$ and a pair of  simply and doubly charged Higgs bosons $H^\pm$ and $H^{\pm\pm}$.
The masse of these Higgs bosons are given by :
\begin{eqnarray}
&& m_{h^0,H^0} = \frac{1}{2}(A+C \pm \sqrt{(A-C)^2 + 4 B^2}) \nonumber\label{eq:mh0mHH}\\
&& m_{H^{\pm\pm}}^2 = \frac{\sqrt{2}\mu{\upsilon_d^2}-\lambda_4\upsilon_d^2\upsilon_t-2\lambda_3\upsilon_t^3}{2v_t}\nonumber\label{eq:mHpmpm}\\
&& m_{H^{\pm}}^2 = \frac{(\upsilon_d^2+2\upsilon_t^2)\,[2\sqrt{2}\mu-\lambda_4\upsilon_t]}{4\upsilon_t}\nonumber\label{eq:mHpm}\\
&& m_{A^0}^2 = \frac{\mu(\upsilon_d^2+4\upsilon_t^2)}{\sqrt{2}\upsilon_t}\label{eq:mA0}\nonumber
\end{eqnarray}
\noindent
The coefficients $A, B$ and $C$ are the entries of the $CP_{even}$ mass matrix defined by,
\begin{eqnarray}
A=\frac{\lambda}{2}v_d^2 ,\quad B=v_d(-\sqrt{2}\mu+(\lambda_1+\lambda_4)v_t)\quad {\rm and} \quad C=\frac{\sqrt{2}\mu\,v_d^2+4(\lambda_2+\lambda_3)v_t^3}{2v_t}
\end{eqnarray}

\section{Theoretical and Experimental Constraints}
The HTM Higgs potential parameters have to obey several constraints originating from theoretical and experimental requirements.\\
{\sl \underline{Boundedness From Below (BFB)}:}
\begin{eqnarray}
&& \lambda \geq 0\,,\,\lambda_2+\lambda_3 \geq 0\,,\,\lambda_1+\lambda_4 \sqrt{\lambda(\lambda_2+\lambda_3)} \geq 0 \,,\, \lambda_1+ \sqrt{\lambda(\lambda_2+\lambda_3)} \geq 0\,,\,\lambda_2+\frac{\lambda_3}{2} \geq 0 \nonumber \\
&& \lambda_3 \sqrt{\lambda} \le |\lambda_4| \sqrt{\lambda_2+\lambda_3} \;\; {\rm or} \;\; 
2 \lambda_1+\lambda_4+\sqrt{(\lambda-\lambda_4^2) (2\frac{\lambda_2}{\lambda_3} + 1)} \geq 0
\end{eqnarray}
{\sl \underline{Unitarity}:}
\begin{eqnarray}
&&|\lambda_1 + \lambda_4| \leq 8 \pi\,,\,\,|\lambda_1| \leq 8 \pi\,,\,\,|\lambda| \leq  16 \pi\,,\,\, |2 \lambda_1 + 3 \lambda_4| \leq 16 \pi\,,\,\,|\lambda_2 + \lambda_3| \leq  4 \pi\,,\,\,|\lambda_2| \leq  4 \pi \nonumber\\
&&| 3 \lambda + 16 \lambda_2 + 12 \lambda_3 \pm \sqrt{(3 \lambda - 16 \lambda_2 - 12 \lambda_3)^2
+ 24 (2 \lambda_1 +\lambda_4)^2} \;| \leq  32 \pi \\
&&|\lambda + 4 \lambda_2 + 8 \lambda_3 \pm \sqrt{(\lambda - 4 \lambda_2 - 8 \lambda_3)^2
+ 16 \lambda_4^2} \;| \leq  32 \pi \,,\,\, |2 \lambda_1 - \lambda_4| \leq 16 \pi\,,\,\,|2 \lambda_2 - \lambda_3| \leq  8 \pi  \nonumber
\end{eqnarray}
{\sl \underline{Bounds on Higgs bosons masses  }:} \\
From the LEP direct search results, the lower bounds on $m_{A^0, H^0} >  80-90$ ~GeV for models with more than one doublet in the case of the neutral scalars.  For the singly charged Higgs mass we use the LEP II latest bounds, $m_{H^{\pm}} \geq 78$~GeV from direct search results, whereas the indirect limit is slightly higher $m_{H^{\pm}} \geq 125$~GeV \cite{lep}.\\
 In the case of the doubly charged Higgs masses, the most recent experimental lower limits  reported by ATLAS and CMS  are respectively $m_{H^{\pm \pm}} \geq 409$~GeV \cite{atlas_dcharged} and $m_{H^{\pm \pm}} \geq 445$~GeV \cite{cms_dcharged} (assuming $Br(H^{\pm \pm}  \to l^{\pm} l^{\pm} =100 \%$). However one can still find scenarios where $m_{H^{\pm \pm}}$ may goes down  to about $100 $ ~GeV.

{\sl \underline{$\rho$ parameter}:} \\
In addition using the $\rho$ parameter in HTM, 
$\rho \simeq 1 - 2 \frac{v_t^2}{v_d^2}$ which indicates a deviation from unity, and imposing consistency with the current limit on $\rho$ from precision measurements \cite{db12}, the upper limit  on $v_t$ is found to be about  about $\leq 5$~GeV. \\

\section{Veltman Conditions}
The Veltman condition \cite{Veltman81} implies that the quadratic divergencies of the two possible tadpoles
$T_{h^0}$ and $T_{H^0}$ of the $h^0$ and $H^0$ CP-even neutral scalar fields vanish. 
It turns out that the two orthogonal combinations 
$T_d=s_{\alpha} T_{h^0} + c_{\alpha} T_{H^0}$ and $T_t=c_{\alpha} T_{h^0} - s_{\alpha} T_{H^0}$ induce simplification.  
Here, $s_{\alpha}$ and $c_{\alpha}$ denote the shortthand notation of cosine and sine for the rotation angle in the $CP_{even}$ Higgs sector. One ends up with the short expressions :
\begin{equation}
T_t =  4 \frac{m_W^2}{v_{sm}^2} (\frac{1}{c_w^2} +1 ) + ( 2 \lambda_1 + 8 \lambda_2 +6 \lambda_3+ \lambda_4) \nonumber
\end{equation}
and
\begin{eqnarray}
T_d &=& - 2 Tr(I_n) \Sigma_f \frac{m_f^2}{v_d^2} + 3 (\lambda + 2 \lambda_1 + \lambda_4) + 2 \frac{m_W^2}{v_{sm}^2} \times ( \frac{1}{c_w^2} +2 )
\end{eqnarray}
corresponding to the tadpoles of the triplet and doublet before symmetry breaking.

\section{Phenomenology}
\begin{figure}
\hspace{-.2in}
\begin{tabular}{ccc} 
{\includegraphics[height=12em]{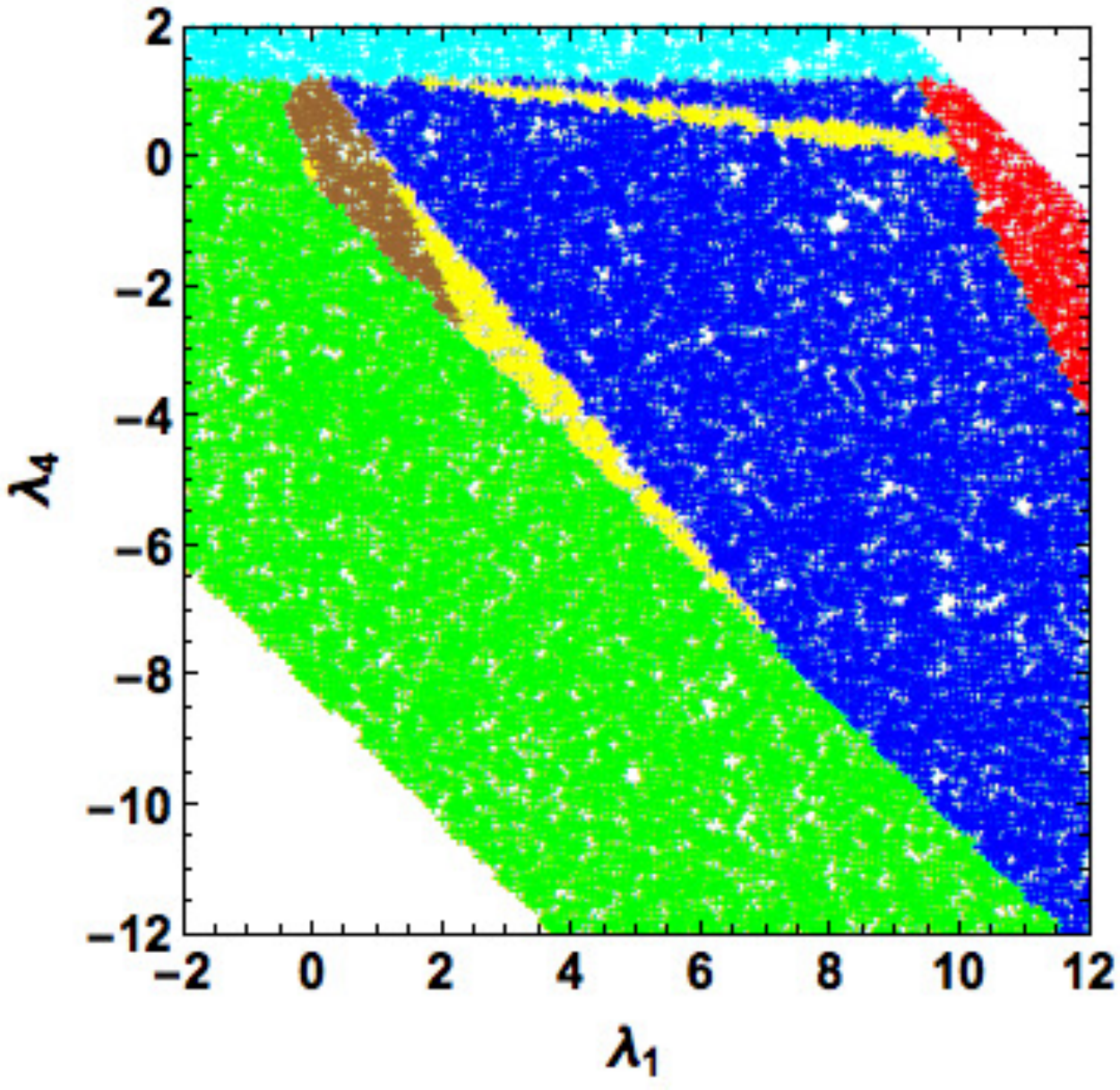}} & {\includegraphics[height=12em]{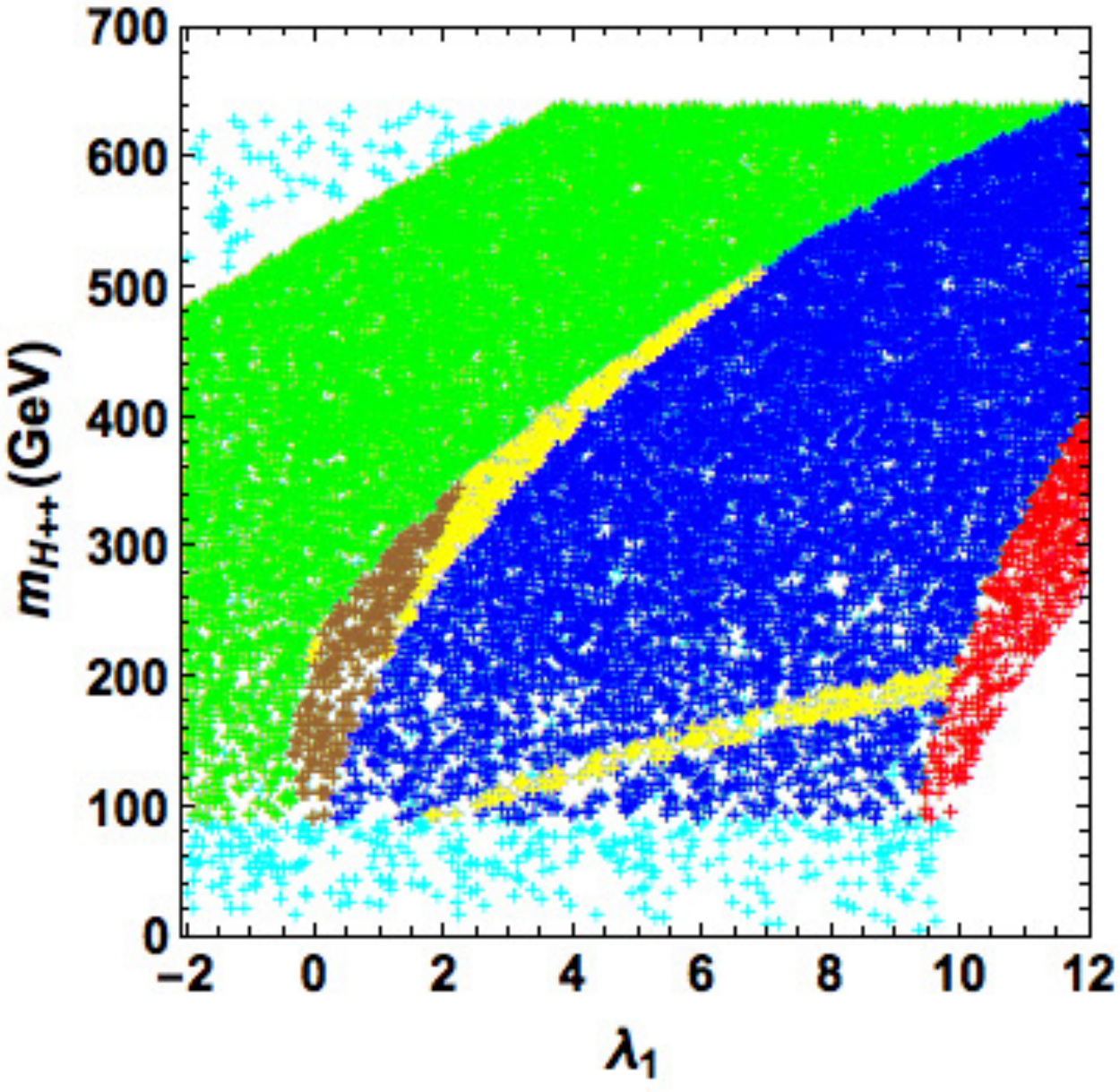}} & {\includegraphics[height=12em]{3.pdf}}
\end{tabular}
\caption{The allowed regions in ($\lambda_{1},\lambda_{4}$) and ($\lambda_{1},m_{H^{\pm\pm}}$)
plans. (\textcolor{cyan}{cyan}) : Excluded 
by $\mu$ constraints, (\textcolor{red}{red}) : Excluded by $\mu$+Unitarity constraints, 
(\textcolor{green}{green}) : Excluded by $\mu$+Unitarity+BFB constraints, (\textcolor{blue}{blue}) : 
Excluded by $\mu$+Unitarity+BFB+$R_{\gamma\gamma}$ constraints, (\textcolor{yellow}{yellow}) : 
Excluded by $\mu$+Unitarity+BFB $R_{\gamma\gamma}$\& $T_d=0$ $\land$ $T_t=0$ constraints. 
Only the brown area obeys  ALL constraints. Our inputs are $\lambda = 0.52$, 
$-2 \le \lambda_1 \le 12$, $\lambda_2 = -\frac{1}{6}$, $\lambda_3 = \frac{3}{8}$, 
$-12 \le \lambda_4 \le 2$, $v_t = 1$ GeV and $\mu = 1$ GeV.}
\label{fig.fig1}
\end{figure}

As a result, we can show that the Veltman condition for HTM field leads  to a drastic reduction of the  parameter space to a relatively small
allowed region (marked in brown in the figures), as one can see in Fig.\ref{fig.fig1}, from which $\lambda_1 \in [-0.3, 2.3]$ corresponding to $m_{H^{\pm \pm}}$ varying from $90$ to $351$ GeV, for which $\lambda_4$ is limited to lie in a reduced interval between $-2.6$ and $1.18$. There are translated to dramatically reduced ranges for the heavy Higgs masses. \\ 

\begin{figure}
\hspace{-.2in}
\begin{tabular}{ccc}
\includegraphics[height=12em]{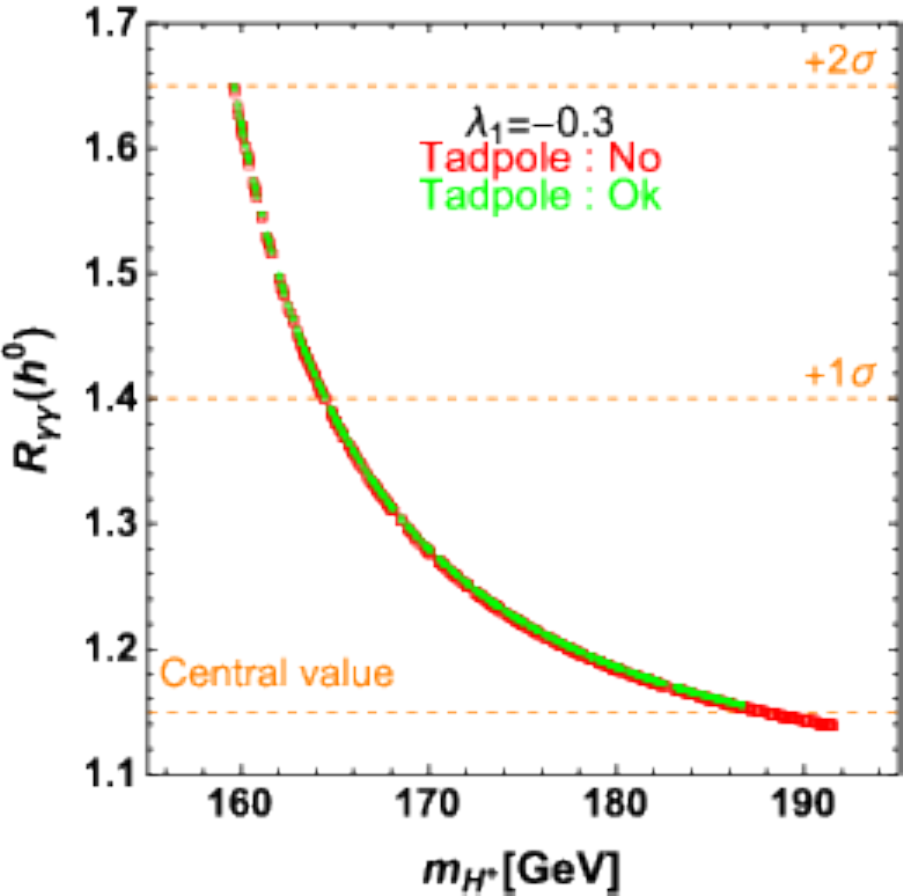} & \includegraphics[height=12em]{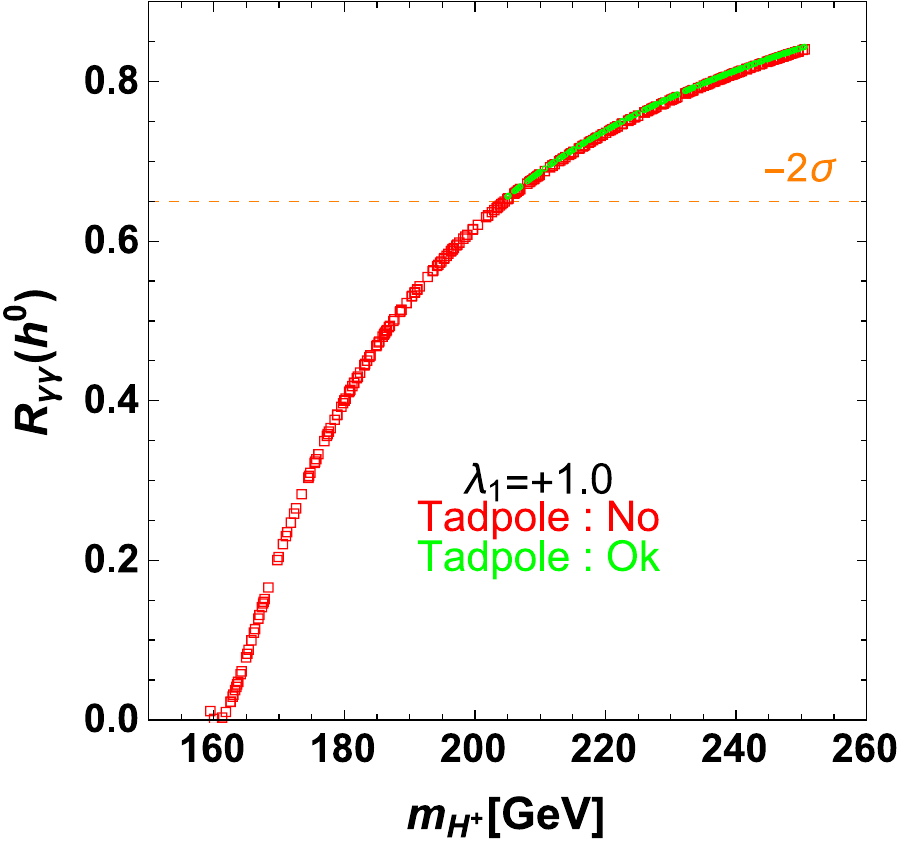} & \includegraphics[height=12em]{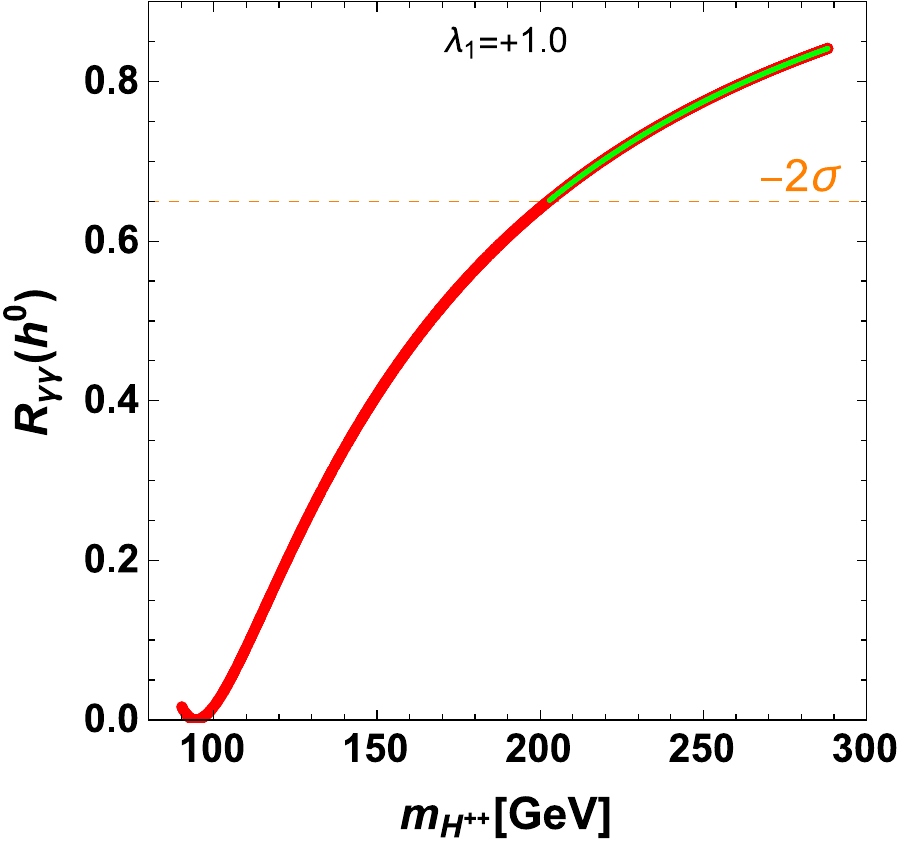}
\end{tabular}
\caption{$R_{\gamma\gamma}(h^0)$ as a function of $m_{H^{\pm}}$ and $m_{H^{\pm\pm}}$ for two values of $\lambda_1$,  
with and without VC.}
\label{fig.fig2}
\end{figure}

This remarkable feature of the effects of the modified VC on the doubly charged Higgs mass is clearly indicated in Fig.\ref{fig.fig2}, where the $R_{\gamma\gamma}$ \cite{aa12} signal strength is plotted as a function of the doubly charged Higgs mass $m_{H^{\pm\pm}}$ for different values of $\lambda_1$.  To close, we have shown that the mVC conditions severly constrain the HTM parameter space. As a byproduct, the  masses of the heavy Higgs bosons are affected and their variation are reduced to lie within the ranges summarised in the Table hereafter.

\begin{table}[!h]
\begin{center}
\renewcommand{\arraystretch}{1.5}
\begin{tabular}{|p{1.10cm}|p{3.3cm}|p{3.cm}|p{3.cm}|p{3.cm}|}
\hline
\hline
$m_{\Phi}$ &  {\bf Unitarity} & {\bf Unitarity + \vspace{-0.1cm}\newline\vspace{-0.1cm}\hspace{-0.1cm} BFB}  &  {\bf Unitarity +\vspace{-0.1cm}\newline\vspace{-0.1cm} \hspace{-0.1cm}BFB +\vspace{-0.1cm}\newline\vspace{-0.1cm} \hspace{-0.1cm}$R_{\gamma\gamma}$} &  {\bf Unitarity +\vspace{-0.1cm}\newline\vspace{-0.1cm} \hspace{-0.1cm}BFB +\vspace{-0.1cm}\newline\vspace{-0.1cm} \hspace{-0.1cm}$R_{\gamma\gamma}$ +\vspace{-0.1cm}\newline\vspace{-0.1cm} \hspace{-0.1cm}mVC} \\
\hline
\hline
$H^0$ &  $[206.8-207.3]$ GeV &  $[206.8-207]$ GeV  &  $[206.8-207]$ GeV  &  $206.8$ GeV  \\ 
\hline
\hline
$A^0$ &  $206.8$ GeV &  $206.8$ GeV &  $206.8$ GeV &  $206.8$ GeV  \\ 
\hline
\hline
$H^\pm$ & $[160-474]$ GeV  &  $[160-474]$ GeV  &  $[160-392]$ GeV  &  $[161-288]$ GeV  \\ 
\hline
\hline
$H^{\pm\pm}$ &  $[90-637]$ GeV  &  $[90-637]$ GeV   &  $[90-513]$ GeV   &  $[90-351]$ GeV   \\ 
\hline
\end{tabular}
\end{center}
\caption{Higgs bosons masses allowed  intervals in the Higgs triplet model resulting from various constraints, including the modified Veltman conditions}
\label{table}
\end{table}

\end{document}